\documentclass{PoS}
\usepackage{epsfig}
\usepackage{graphicx}
\newcommand{\be}{\begin{eqnarray}}
\newcommand{\ee}{\end{eqnarray}}
\newcommand{\beq}{\begin{equation}}
 \newcommand{\eeq}{\end{equation}}
\title{Heavy flavour production in $pp$ collisions and intrinsic quark components
in proton
}

\ShortTitle{Heavy flavour production in $pp$ collisions}

\author{\speaker{G.~I.~Lykasov} \\
        Joint Institute for Nuclear Research\\
        E-mail: \email{lykasov@jinr.ru}}

\author{A.~A.~Grinyuk\\
       Joint Institute for Nuclear Research\\
       E-mail: \email{andrei.grinyuk@gmail.com}}
\author{ I.~V.~Bednyakov\\
        Joint Institute for Nuclear Research\\
        E-mail: {bednyakovi@gmail.com}}

\abstract{The LHC data on the forward heavy flavour hadron production can be a new unique source for 
estimation of intrinsic heavy quark contributions to the proton. 
We discuss in detail the $D$-meson production in $pp$ collisions at the LHC
including the intrinsic charm in the proton.
We present also some predictions for the $K$-meson production in the $pp$ collision 
at the initial energies 158 GeV and 7 TeV made within 
the perturbative QCD including the intrinsic strangeness in the proton that can 
be verified in the NA61 experiment and at LHC.
}	

\FullConference{XXI International Baldin Seminar on High Energy Physics Problems\\
		 September 10 -15, 2012\\
		 JINR, Dubna, Russia}

\begin{document}
\section{Intrinsic heavy flavours in the proton} 
      The Large Hadron Collider (LHC) at CERN can be a useful laboratory for 
      investigation of the unique structure of the proton, in particular for 
      the study of the parton distribution functions (PDFs) with high accuracy.
      It is well known that the precise knowledge of these PDFs is very important 
      for verification of the Standard Model and search for New Physics. 

      By definition, the PDF $f_a(x,\mu)$ is a function of the proton momentum fraction $x$ 
      carried by parton $a$ (quark $q$ or gluon $g$) at the QCD momentum transfer scale $\mu$. 
      For small values of $\mu$, corresponding to the long distance scales less than $1/\mu_0$, 
      the PDF cannot be calculated from the first principles of QCD 
      (although some progress in this direction 
      has been recently achieved within the lattice methods 
\cite{LATTICE}). 
      The PDF $f_a(x,\mu)$ at $\mu>\mu_0$ can be calculated by 
      solving the perturbative QCD evolution equations (DGLAP) 
\cite{DGLAP}.  
      The unknown (input for the evolution) functions $f_a(x,\mu_0)$ 
      can usually be found empirically from some 
      ``QCD global analysis'' 
\cite{QCD_anal1,QCD_anal2} of a large variety of data, typically at $\mu>\mu_0$. 

     In general, almost all $pp$ processes that took place at the LHC energies, 
     including the Higgs boson production,
     are sensitive to the charm $f_c(x,\mu)$ or bottom $f_b(x,\mu)$ PDFs. 
     Nevertheless, within the global analysis 
     the charm content of the proton at $\mu\sim\mu_c$ and 
     the bottom one at $\mu\sim\mu_b$ are both assumed to be negligible.
     Here $\mu_c$ and $\mu_b$ are typical energy scales relevant to the 
     $c$- and $b$-quark 
     QCD excitation in the proton.
     These heavy quark components arise in the proton only perturbatively
     with increasing $Q^2$-scale 
     through the gluon splitting in the DGLAP $Q^2$ evolution 
\cite{DGLAP}. 
     Direct measurement of the open charm and open bottom production 
     in the deep inelastic processes (DIS) confirms the perturbative 
     origin of heavy quark flavours 
\cite{H1:2005}. 
     However, the description of these experimental data is 
     not sensitive to the heavy quark distributions at relatively 
     large $x$ ($x>0.1$). 

\begin{figure}[h]
\centerline{\includegraphics[width=0.50\textwidth]{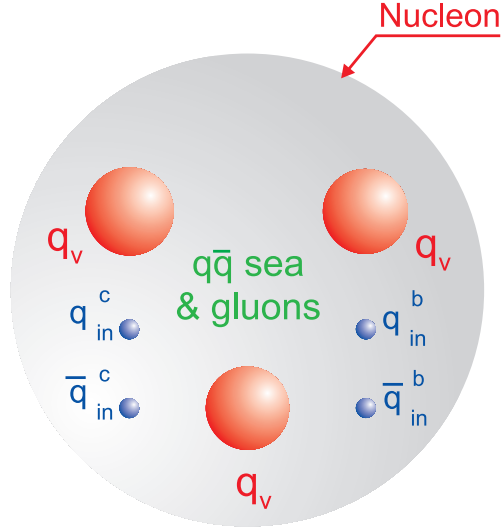}}
\caption{Schematic presentation of a nucleon consisting of three valence quarks
         q$_{\rm v}$, quark-antiquark q${\bar {\rm q}}$ and gluon sea, 
	 and pairs of the intrinsic charm 
	 (q$_{\rm in}^{\rm c}{\bar {\rm q}_{\rm in}^{\rm c}}$) 
	 and intrinsic bottom quarks 
	 (q$_{\rm in}^{\rm b}{\bar {\rm q}_{\rm in}^{\rm b}}$). 
}
\label{Fig_IQ}
\end{figure}

     As was assumed by Brodsky with coauthors in 
\cite{Brodsky:1980pb, Brodsky:1981}, 
     there are {\it extrinsic} and {\it intrinsic} 
     contributions to the quark-gluon structure of the proton. 
     {\it Extrinsic} (or ordinary) quarks and gluons are generated on 
     a short time scale associated with a large-transverse-momentum processes.
     Their distribution functions satisfy the standard QCD evolution equations. 
     {\it Intrinsic} quarks and gluons exist
     over a time scale which is independent of any probe momentum transfer. 
     They can be associated with bound-state 
{(zero-momentum transfer regime)} hadron dynamics and 
     are believed to be of nonperturarbative origin.
Figure~\ref{Fig_IQ}  gives 
     a schematic view 
     of a nucleon, which consists of three valence quarks q$_{\rm v}$, 
     quark-antiquark q${\bar {\rm q}}$ and gluon sea, and, for example,  
     pairs of the {\it intrinsic} charm 
(q$_{\rm in}^{\rm c}{\bar {\rm q}_{\rm in}^{\rm c}}$) and 
     {\it intrinsic} bottom quarks 
(q$_{\rm in}^{\rm b}{\bar {\rm q}_{\rm in}^{\rm b}}$). 

       It was shown in 
\cite{Brodsky:1981}
       that the existence of {\it intrinsic} heavy quark pairs 
       $c{\bar c}$ and $b{\bar b}$ within the proton state 
       could be due to the virtue of gluon-exchange and vacuum-polarization graphs. 
       On this basis, 
       within the MIT bag model 
\cite{Golowich:1981}, 
       the probability to find the five-quark component 
       $|uudc{\bar c}\rangle$ bound within the nucleon bag 
       was estimated to be about 1--2\%. 

       Initially in 
\cite{ Brodsky:1980pb,Brodsky:1981}
       S.Brodsky with coauthors 
       have proposed 
       existence of the 5-quark state $|uudc{\bar c}\rangle$ 
       in the proton 
(Fig.~\ref{Fig_IQ}). 
       Later some other models were developed. 
       One of them considered a quasi-two-body state 
       ${\bar D}^0(u{\bar c})\, {\bar\Lambda}_c^+(udc)$ in the proton 
\cite{Pumplin:2005yf}. 
       In 
\cite{Pumplin:2005yf}--\cite{Nadolsky:2008zw} 
       the probability to find the intrinsic charm (IC) in the proton 
       (the weight of the relevant Fock state in the proton)
       was assumed to be 1--3.5\%. 
       The probability of the intrinsic bottom (IB) in the proton 
       is suppressed by the factor $m^2_c/m^2_b\simeq 0.1$ 
\cite{Polyakov:1998rb}, where $m_c$ and $m_b$ are the masses of 
       the charmed and bottom quarks. 
       Nevertheless, it was 
       shown that the IC 
       could result in a sizable contribution 
       to the forward charmed meson production
\cite{Goncalves:2008sw}. 
       Furthermore the IC ``signal'' can 
       constitute almost 
       100\% 
       of the inclusive spectrum of $D$-mesons produced at 
       high pseudorapidities $\eta$ 
       and large transverse momenta $p_T$
       in $pp$ collisions at LHC energies 
\cite{LBPZ:2012}. 

       If the distributions of the intrinsic charm or bottom in the 
       proton are hard enough and are similar in the shape to the valence quark distributions
       (have the valence-like form),  
       then the production of the charmed (bottom) mesons or charmed (bottom) 
       baryons in the fragmentation region should be similar 
       to the production of pions or nucleons. 
       However, the yield of this production depends on the probability to find the 
       intrinsic charm or bottom in the proton, but this yield looks too small.     
       The PDF which included the IC contribution in the proton 
       have already been used in the perturbative QCD calculations in  
\cite{Pumplin:2005yf}-\cite{Nadolsky:2008zw}.

\begin{figure}[p] 
\begin{center}
\begin{tabular}{cc}
\epsfig{file=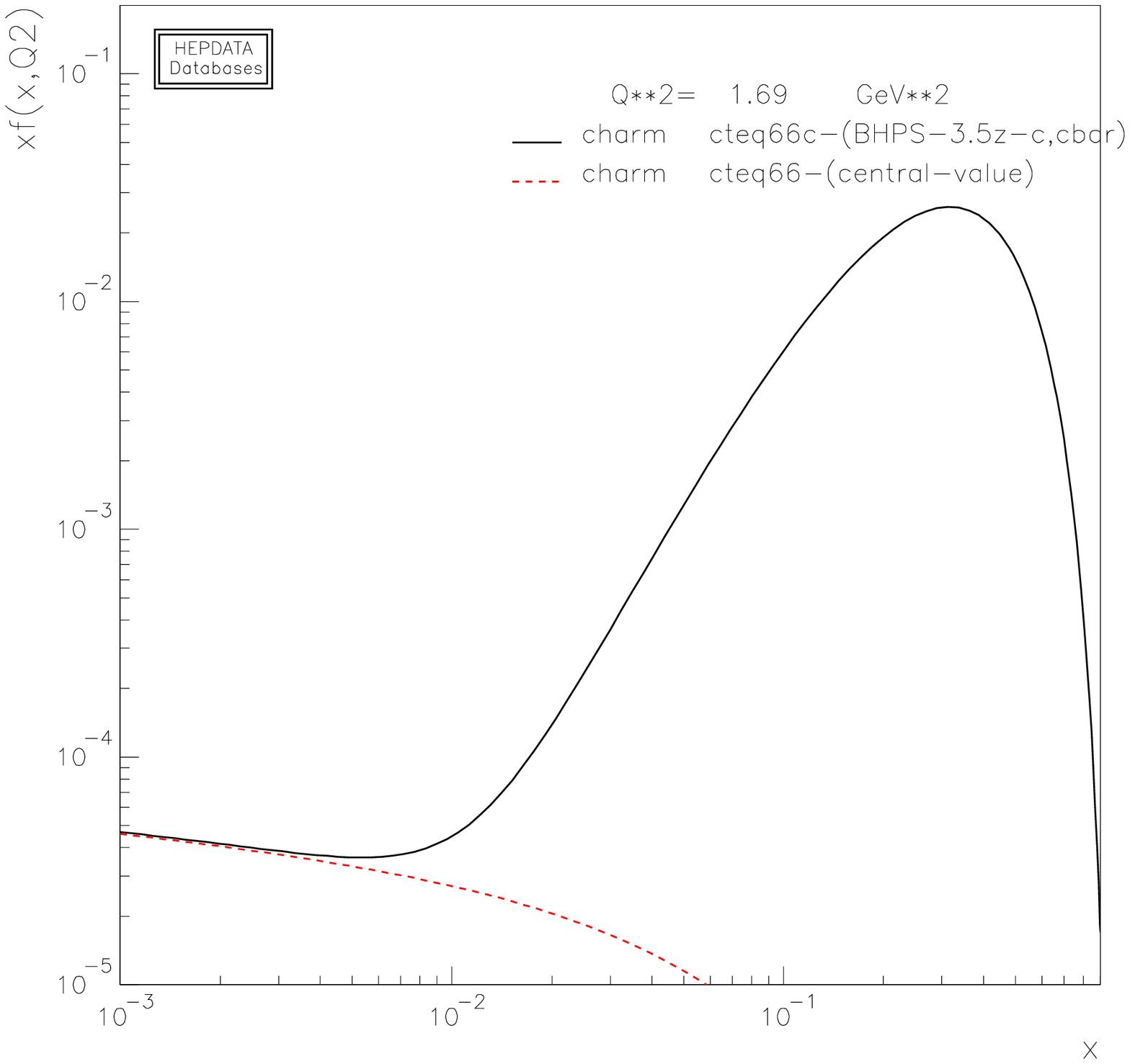,width=0.6\linewidth}
\end{tabular}
\end{center}
 \caption{The distributions of charmed quarks in the proton; the dashed line is the 
sea charmed quarks $c(x)$, the solid curve is the sum of the intrinsic charm
ans the sea one  $c(x)+c_{\rm in}(x)$}.
\label{Fig_2IC}

\begin{center}
\begin{tabular}{cc}
\epsfig{file=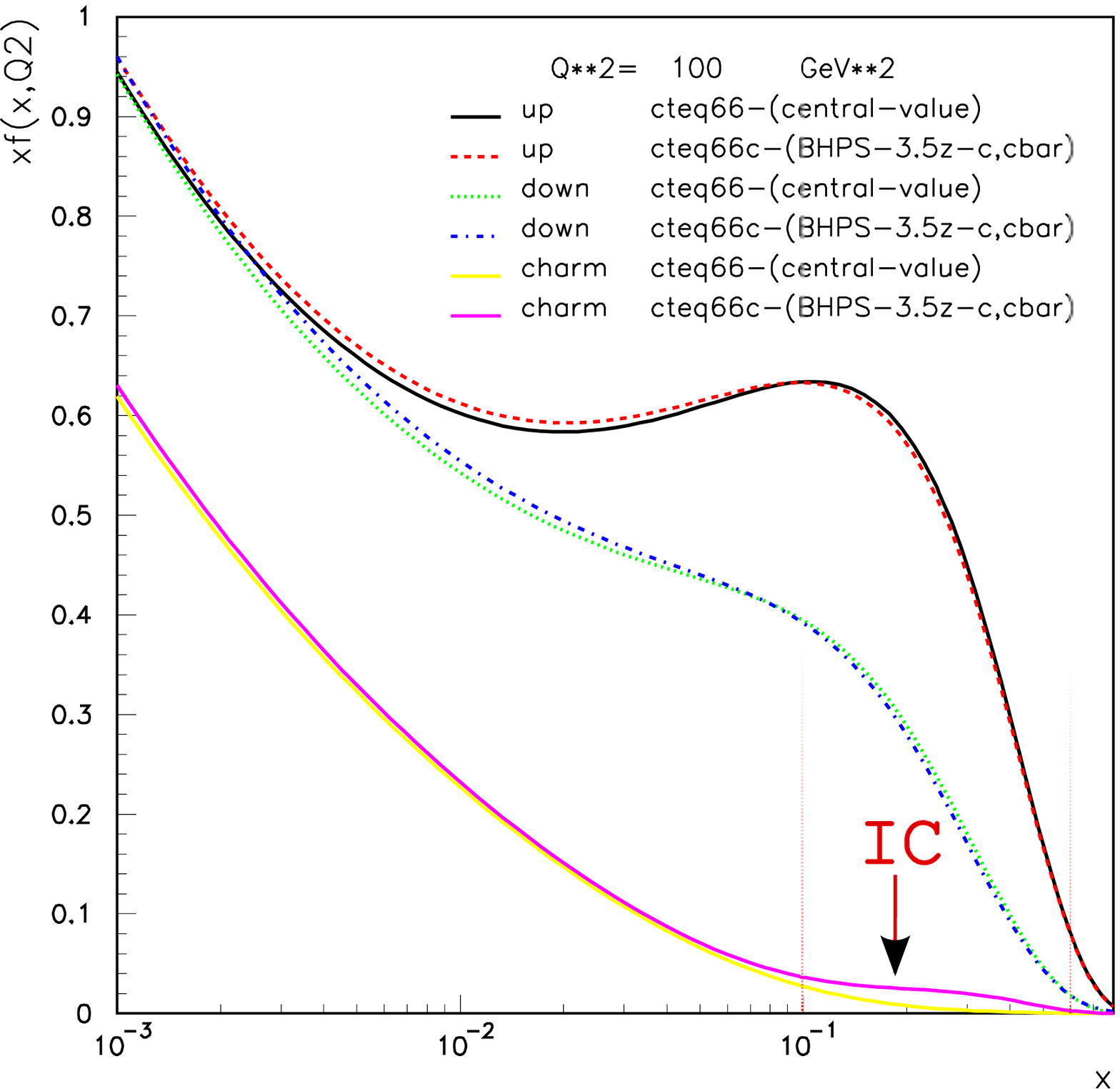,width=0.6\linewidth}
\end{tabular}
\end{center}
 \caption{   
     The distributions of the valence ($u,d$) quarks and sea quarks in the proton
     as a function of the Bjorken variable $x$.}
\label{Fig_pdfs}
\end{figure}

       The probability distribution for the 5-quark state ($uudc{\bar c}$) 
       in the light-cone description of the proton was first calculated in 
\cite{Brodsky:1980pb}. 
       Assuming that the light quark ($u,d$) masses and the proton
       mass are smaller than the $c$-quark mass one can get the following form 
       for this probability
\cite{Pumplin:2007wg} at $Q^2=m_c^2$ ($m_c=1.69$ is the c-quark mass):
\begin{eqnarray}
\frac{dP}{dx}=f_c(x)=f_{\bar c}(x)={\cal N}x^2 \, 
\left\{(1-x)(1+10x+x^2)+6x(1+x)\ln(x)\right\}~,
\label{def:fcPumpl}
\end{eqnarray} 
     where the normalization constant ${\cal N}$ determines some 
     probability $w^{}_{\rm IC}$ to find 
     the Fock state $|uudc{\bar c}\rangle$ in the proton. 
     The solid line in       
Fig.~\ref{Fig_2IC} 
     shows the intrinsic charm PDF $xf_c(x)$ as a function of $x$ at $Q^2=m_c^2$
     when the probability $w^{}_{\rm IC}= 3.5$\%. 
     The dashed curve in
Fig.~\ref{Fig_2IC} is the density distribution of the (ordinary) sea charm in the proton.    
      One can see from 
Fig.~\ref{Fig_2IC} 
      that the IC distribution (with $w^{}_{\rm IC} = 3.5$\% 
\cite{Nadolsky:2008zw}) given by Eq. 
(\ref{def:fcPumpl}) 
      has rather visible 
      enhancement at $x\sim 0.2-0.3$ and 
      it is much larger (a few orders of magnitude) 
      than the sea (ordinary) charm density distribution in the proton. 
    The valence quark density distributions ($xf_{u.d}$), 
    the distributions of the sea charm ($c_{\rm sea}$) and the intrinsic charm 
    ($xf_{\rm IC}(x)$) at $Q^2=100$~GeV/$c^2$ are presented in 
Fig.~\ref{Fig_pdfs}.
     One can see from 
Fig.~\ref{Fig_pdfs}.
     that the sea charm and the intrinsic charm distributions are much ``smaller''
     than the valence ones in the whole region of $x$. However, in the hard $pp$ 
     collisions the gluons and sea quarks make the main contribution to the 
     inclusive hadron spectra. Therefore, the inclusion of the intrinsic heavy quark 
     components in the proton makes sense.  

   Due to the nonperturbative {\it intrinsic} heavy quark components one can expect 
   some excess of the heavy quark PDFs over 
   the ordinary sea quark PDFs at $x>0.1$. 
   The ``signal'' of these components can be visible in the observables of 
   the heavy flavour production in semi-inclusive $ep$ DIS and inclusive 
   $pp$ collisions at high energies. 
   For example, it was recently shown that rather good description of the HERMES data 
   on the $xf_s(x,Q^2)+xf_{\bar s}(x,Q^2)$ at $x>0.1$ and 
   $Q^2=2.5$~GeV$/c^2$
\cite{IS:2012} could be achieved due to existence of intrinsic strangeness in the proton.  
   Similarly, possible existence of the intrinsic charm in the proton
   can lead to some enhancement in the inclusive spectra of the open charm hadrons, 
   in particular $D$-mesons, produced at the LHC in $pp$-collisions 
   at high pseudorapidities $\eta$
   and large transverse momenta $p_T$ 
\cite{LBPZ:2012}.

\section{Intrinsic heavy quarks at the LHC}
$\bullet~${\bf Intrinsic charm}\\
    It is known that 
    in the open charm/beauty $pp$-production at large momentum transfer 
    the hard QCD interactions of two sea quarks, two gluons and 
    a gluon with a sea quark play the main role.
According to the model of hard scattering 
\cite{AVEF:1974}--\cite{FF:AKK08}
the relativistic invariant
inclusive spectrum of the hard process $p+p\rightarrow h+X$ can be related to
the elastic parton-parton subprocess $i+j\rightarrow i^\prime +j^\prime$,
where $i,j$ are the partons (quarks and gluons). 
This spectrum can be presented in the following general form 
\cite{FF}--\cite{FFF2} (see also \cite{BGLP:2011,BGLP:2012}):
\begin{eqnarray}
\label{def:rho_c} 
E\frac{d\sigma}{d^3p}= 
\sum_{i,j}\!\int\! d^2k_{iT}\!\int\! d^2k_{jT}\!\int_{x_i^{\min}}^1dx_i\!\int_{x_j^{\min}}^1dx_j
f_i(x_i,k_{iT})f_j(x_j,k_{jT}) 
\frac{d\sigma_{ij}({\hat s},{\hat t})}{d{\hat t}}\frac{D_{i,j}^h(z_h)}{\pi z_h}.
\label{def:hscm} 
\end{eqnarray}
   Here $k_{i,j}$ and $k_{i,j}^\prime$ are the four-momenta of the partons $i$ or $j$ 
   before and after the elastic parton-parton scattering, respectively; 
   $k_{iT}, k_{jT}$ are the transverse momenta of the partons $i$ and $j$;  
   $z$ is the fraction of the hadron momentum from the parton momentum; 
   $f_{i,j}$ is the PDF; and $D_{i,j}$ is the fragmentation function (FF) 
   of the parton $i$ or $j$ into a hadron $h$.

    When 
    the transverse momenta of the partons are neglected 
    in comparison with the longitudinal momenta, 
    the variables ${\hat s}$, ${\hat t}$, ${\hat u}$ and $z_h$ can be 
    presented in the following forms \cite{FF}:
\be
{\hat s}=x_i x_j s,\quad {\hat t}=x_i \frac{t}{z_h}, \quad
{\hat u}=x_j \frac{u}{z_h},\quad z_h=\frac{x_1}{x_i}+\frac{x_2}{x_j},
\label{def:stuzh}
\ee
     where
\be
x_1=-\frac{u}{s}=\frac{x_T}{2}\cot({\theta}/{2}), \quad
x_2=-\frac{t}{s}=\frac{x_T}{2}\tan({\theta}/{2}), \quad
x_T=2\sqrt{t u}/s=2p_T/\sqrt{s}.
\ee
      Here as usual, 
      $s=(p_1+p_2)^2$,
      $t=(p_1-p_1^\prime)^2$,
      $u=(p_2-p_1^\prime)^2$, 
     and $p_1$, $p_2$, $p_1^\prime$ are the 4-momenta of the colliding protons 
     and the produced hadron $h$, respectively; 
     $\theta$ is the scattering angle for the hadron $h$ in the $pp$ c.m.s.
     The lower limits of the integration in
(\ref{def:rho_c}) are 
\be
x_i^{\min}=\frac{x_T \cot(\frac{\theta}{2})}{2-x_T \tan(\frac{\theta}{2})}, \qquad
x_j^{\min}=\frac{x_i x_T \tan(\frac{\theta}{2})}{2x_i-x_T \cot(\frac{\theta}{2})}.
\label{def:xijmn}
\ee 
Actually, the parton distribution functions $f_i(x_i,k_{iT})$ also depend on  
the four-momentum transfer squared $Q^2$ that is related to the Mandelstam variables
${\hat s},{\hat t},{\hat u}$ for the elastic parton-parton scattering \cite{FFF2} 
\be
Q^2~=~\frac{2{\hat s}{\hat t}{\hat u}}{{\hat s}^2+{\hat t}^2+{\hat u}^2}
\label{def:Qsqr}
\ee 

    Calculating spectra by Eq.(\ref{def:rho_c}) we used 
    the PDF which includes the IS (and does not include it) \cite{Nadolsky:2008zw},
     the FF of the type AKK08 
\cite{FF:AKK08}  and 
    $d\sigma_{ij}({\hat s},{\hat t})/d{\hat t}$ 
    calculated within the LO QCD and presented, for example, in 
\cite{Mangano:2010}.  

\begin{figure}[h!]
\centering{
\epsfig{file=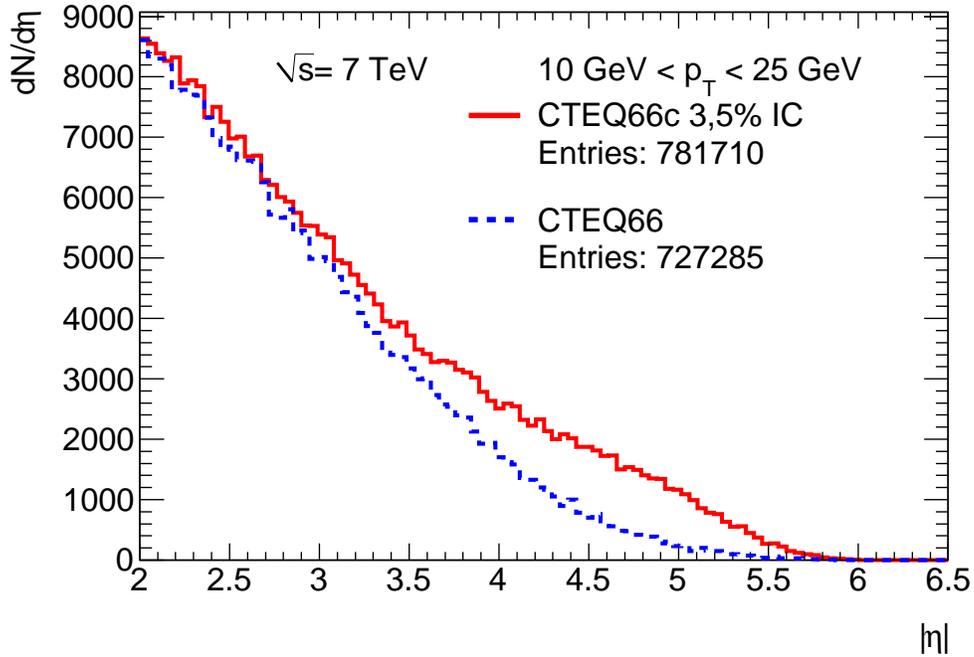,width=0.9\linewidth}}
\caption{
  The $D+{\bar D}_0$ distributions (with and without intrinsic charm contribution) 
  over the pseudorapidity $\eta$ for $pp\rightarrow (D_0+{\bar D}_0) + X$  
  at $\sqrt{s}=7$ TeV and $10\leq p_T\leq 25$ GeV$/c$ \cite{LBPZ:2012}.}
\label{Fig_4}
\end{figure}

    In particular, 
Fig.~\ref{Fig_4} shows our estimation of the inclusive yield 
    of single $D^0$-mesons in $pp$-production
    at $\sqrt{s}=7$ TeV and 10~GeV$/c<p_T<25$~GeV$/c$ 
    as a function of the pseudorapidity $\eta$.
    This estimation is obtained within PYTHIA8, where the
    PDF CTEQ66 (without the IC, given by the dashed blue curve in 
Fig.~\ref{Fig_4})
    and CTEQ66c (with the IC for $w^{}_{\rm IC} \simeq 3.5$\%, given by 
    the solid red curve) were used at $Q^2=m_c^2=$(1.69 GeV)$^2$ 
\cite{Nadolsky:2008zw}.

    One can see from 
Fig.~\ref{Fig_4}
    that there is some enhancement due to the IC contribution at large $\eta$,
    which is due to the 
    above-mentioned enhancement in the IC PDF at $x>0.1$, given in 
Fig.~\ref{Fig_2IC}. 
    Its amount increases with growing $p_T$. 
    For example, due to the IC the spectrum increases by a factor of 2 at 
    $\eta=4.5$.
    A similar  effect was predicted in 
\cite{Kniehl:2012ti}.

     One can see that the Feynman variable $x_F$ of the produced hadron, 
     for example, the $D^0$-meson, can be expressed via  
     the variables $p_T$ and $\eta$, or $\theta$ 
     the hadron scattering angle in the $pp$ c.m.s, 
\begin{eqnarray}
x_F \equiv \frac{2p_{z}}{\sqrt{s}}
=\frac{2p_T}{\sqrt{s}}\frac{1}{\tan\theta}
=\frac{2p_T}{\sqrt{s}}\sinh(\eta). 
\label{def:xFptteta}
\end{eqnarray} 
     At small scattering angles of the produced hadron this formula becomes 
\begin{eqnarray}
x_F\sim \frac{2p_T}{\sqrt{s}}\frac{1}{\theta}.
\label{def:xFtetapt}
\end{eqnarray} 
       It is clear that for fixed $p_T$ an outgoing hadron must possess a very small $\theta$ or very large $\eta$
        in order to have large $x_F$ (to follow forward, or backward direction).

       In the fragmentation region (of large $x_F$) the Feynman variable $x_F$ 
       of the produced hadron is related to 
       the variable $x$ of the intrinsic charm quark in the proton, and  
       according to the longitudinal momentum conservation law, 
       the $x_F \simeq x$ (and $x_F < x$). 
       Therefore, the visible excess of the solid (red) histogram over 
       the dashed (blue) one in 
Fig.~\ref{Fig_4} at $\eta>3.5$ 
       is due to the enhancement of the IC distribution 
(see Fig.~\ref{Fig_2IC}) 
       at $x>$ 0.1.

      One expects similar enhancement in 
      the experimental distributions of the open bottom production 
      due to the (hidden) intrinsic bottom (IB) in the proton, which could have  
      the PDF very similar to the one given in
(\ref{def:fcPumpl}). 
      However, the probability $w_{\rm IB}$ to find in the proton 
      the Fock state with the IB contribution $|uudb{\bar b}\rangle$ 
      is about 10 times lower than the IC probability $w_{\rm IC}$ 
      due to the relation $w_{\rm IB}/w_{\rm IC}\sim m_c^2/m_b^2$, 
      where $m_b$ is the bottom quark mass 
\cite{Brodsky:1981,Polyakov:1998rb}.

$\bullet~${\bf Intrinsic strangeness}\\
Let us analyze now how the possible existence of the intrinsic strangeness in the proton 
can be visible in $pp$ collisions. For example, consider the $K^-$-meson production in 
the process $pp\rightarrow K^-+X$. Considering the intrinsic strangeness in the proton \cite{IS:2012}
we calculated the inclusive spectrum $ED\sigma/d^3p$
of such mesons within the hard scattering model (Eq.(\ref{def:hscm})), 
which describes satisfactorily the
HERA and HERMES data on the DIS. The FF and the parton cross sections were taken from
\cite{FF:AKK08,Mangano:2010}, respectively, as mentioned above.
\begin{figure}[h!]
\centering{
\epsfig{file=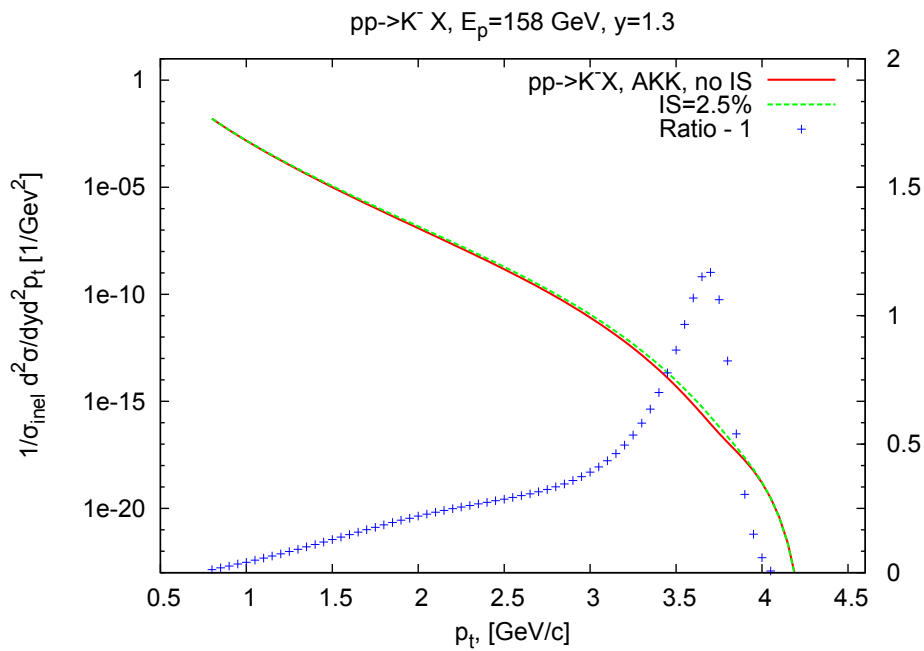,width=0.8\linewidth}}
\caption{
  The $K^-$-meson distributions (with and without intrinsic strangeness contribution) 
  over the transverse momentum $p_t$ for $pp\rightarrow K^- + X$  
  at the initial energy $E=$ 158 GeV, the rapidity $y=$1.3 and $p_t\geq$ 0.8 GeV$/$c.}
\label{Fig_5}
\end{figure} 
   \begin{figure}[h!]
\centering{
\epsfig{file=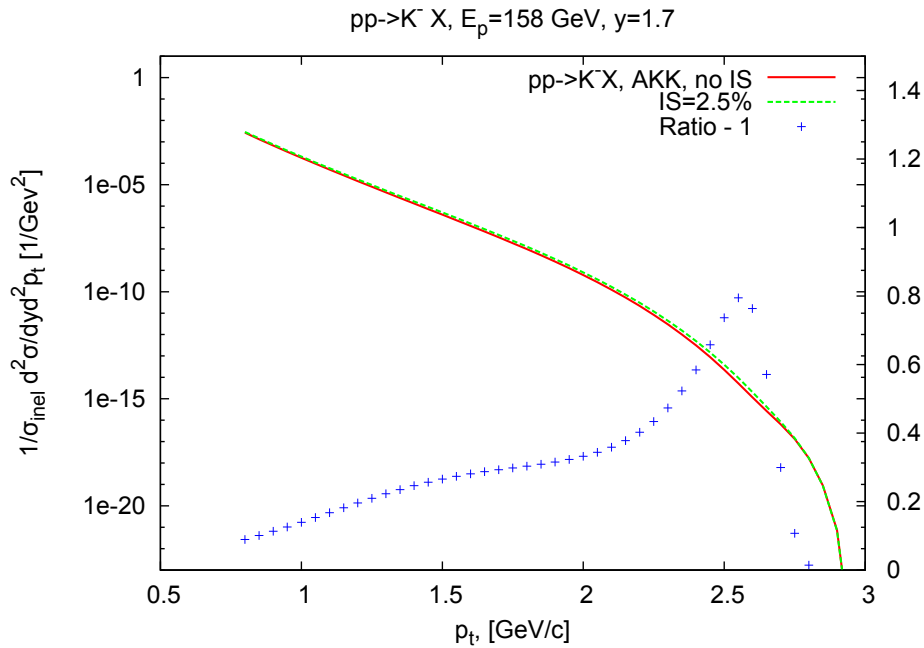,width=0.8\linewidth}}
\caption{
  The $K^-$-meson distributions (with and without intrinsic strangeness contribution) 
  over the transverse momentum $p_t$ for $pp\rightarrow K^- + X$  
  at the initial energy $E_p=$ 158 GeV, the rapidity $y=$1.7 and $p_t\geq$ 0.8 GeV$/$c.}
\label{Fig_6}
\end{figure}  
In Figs.~(\ref{Fig_5},\ref{Fig_6}) the inclusive $p_t$-spectra of $K^-$-mesons produced in $pp$
collision at the initial energy $E_p=$158 GeV are presented at the rapidity $y=$1.3 (Fig~\ref{Fig_5})
and $y=$1.7 (Fig~\ref{Fig_6}). The solid lines in  Figs.~(\ref{Fig_5},\ref{Fig_6}) correspond to our
calculation ignoring the {\it intrinsic strangeness} (IS) in the proton and the dashed curves correspond 
to the calculation including the IS with the probability about 2.5$\%$, according to \cite{IS:2012}. 
The crosses show the ratio of our calculation with the IS and without the IS minus 1.
One can see from Figs.~(\ref{Fig_5},\ref{Fig_6}, right axis) that the IS signal can be above 200 $\%$ 
at $y=$ 1.3, $p_t=$ 3.6-3.7 Gev$/$c and slightly smaller, than 200 $\%$ at 
$y=$ 1.7, $p_t\simeq$ 2.5 Gev$/$c. Actually, this is our prediction for the NA61 experiment that
is now under way at CERN.  

\begin{figure}[h!]
\centering{
\epsfig{file=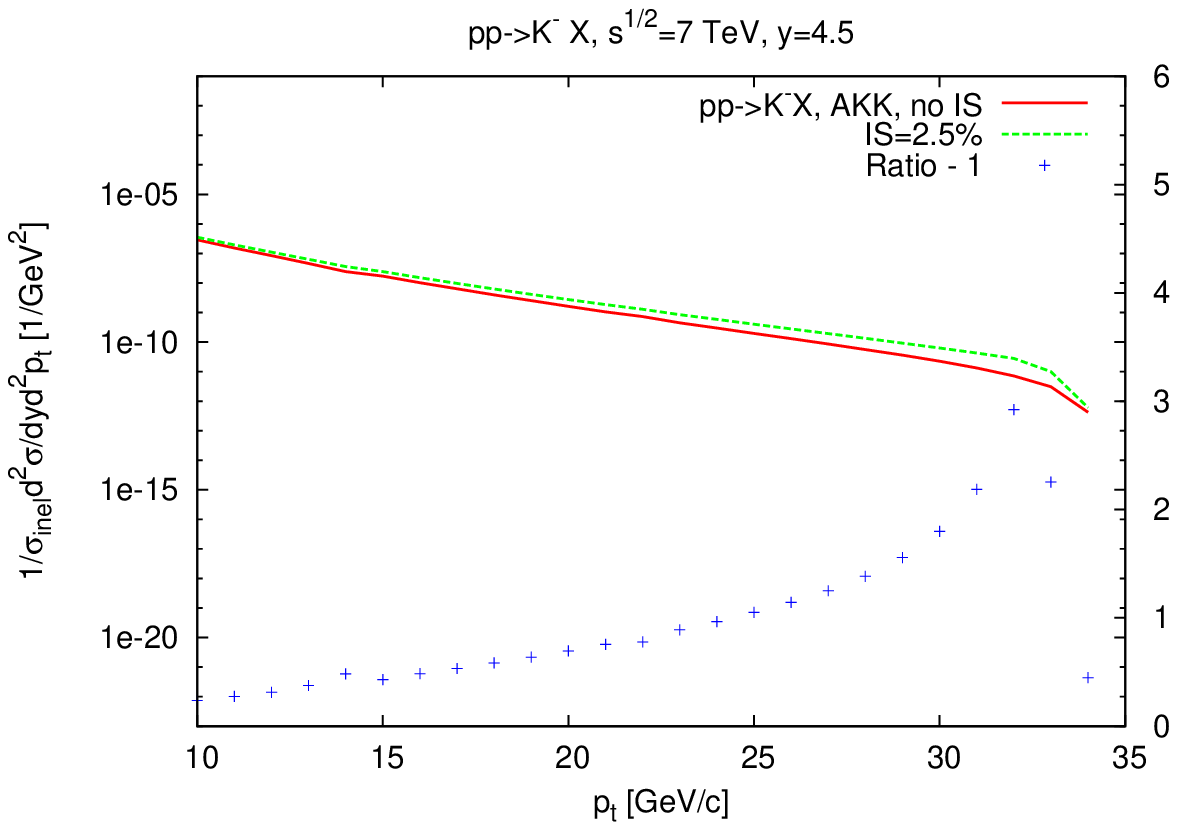,width=0.8\linewidth}}
\caption{
  The $K^-$-meson distributions (with and without intrinsic strangeness contribution) 
  over the transverse momentum $p_t$ for $pp\rightarrow K^- + X$  
  at $\sqrt{s}=$ 7 TeV, the rapidity $y=$4.5 and $p_t\geq$ 10 GeV$/$c.}
\label{Fig_Kmin_7T}
\end{figure}

\begin{figure}[h!]
\centering{
\epsfig{file=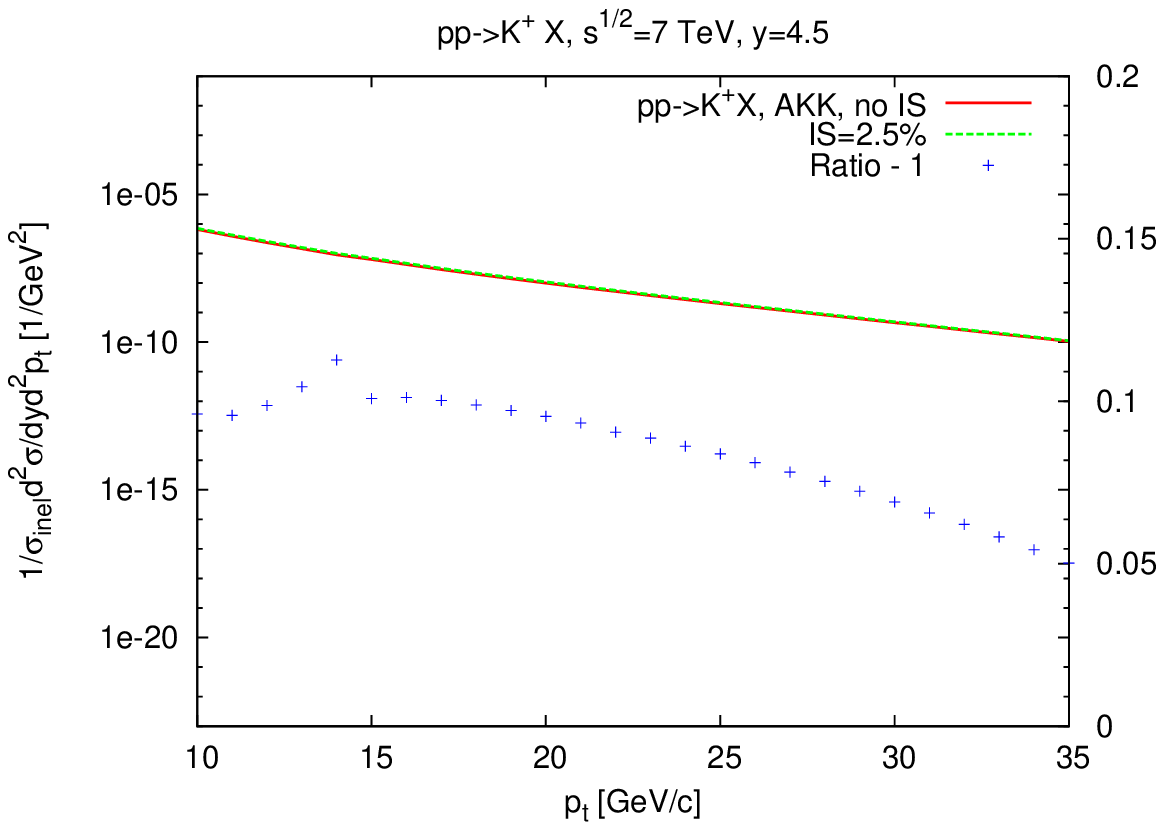,width=0.8\linewidth}}
\caption{
  The $K^+$-meson distributions (with and without intrinsic  strangeness contribution) 
  over the transverse momentum $p_t$ for $pp\rightarrow K^+ + X$  
  at $\sqrt{s}=$ 7 TeV, the rapidity $y=$4.5 and $p_t\geq$ 10 GeV$/$c.}
\label{Fig_Kplus_7T}
\end{figure}

We also calculated the inclusive spectra of $K^-$ and $K^+$ mesons produced in the $pp$ collision
at LHC energies. Note that the spectra of $K^-$-mesons, which consist of $s$- and ${\bar u}$- quarks, 
can give us information on the IS at large $x_F$ ($x_F>$0.1) or at the certain region of $p_T$ and $\eta$, according 
to Eq.(\ref{def:xFptteta}) and Fig.~(\ref{Fig_2IC}), because they are produced mainly from the fragmentation
of the strange sea ($s_{sea}$) and the {\it intrinsic} strange ($s_{in}$) quarks in the proton. 
One can see from Fig.~(\ref{Fig_Kmin_7T}, right axis) that the IS signal at 
$\sqrt{s}=$7 TeV, $y=$4.5 can be about 400 $\%$ at $p_T\simeq $ 32-33 GeV$/$c, the measuring of which at LHC
can be difficult. Anyway, at $p_T>$15 GeV$/$c  and $y=$4.5 the IS signal is about 200 $\%$ and
more, as the crosses show in Fig.~(\ref{Fig_Kmin_7T},right axis). 
However, the spectra of $K^+$-mesons almost do not give us such information because $K^+$-meson consists of
the $u$- and ${\bar s}$-quarks and its production at large $x_F$ is due to the fragmentation of the valence 
$u$-quark by the $pp$ collision. 
It illustrates Fig.~(\ref{Fig_Kplus_7T}), the notations are the same as in  
Figs.~(\ref{Fig_5}-\ref{Fig_Kmin_7T}).  
The similar effect but for the $IC$ signal can be visible in the inclusive spectra of $D^-$- and $D^+$-mesons 
produced in $pp$ collision at LHC energies. 

Let us also note that we calculated the inclusive spectra of $K$- and $D$-mesons within the LO QCD. The NLO calculation,
in principle, can change the inclusive spectra. However, the ratio of the spectrum of $K^-$ (or $D^-$) to the spectrum of $K^+$
(or $D^+$) can not be very sensitive to the NLO corrections.

\section{Conclusion} 
          In this paper we have shown that possible existence of the {\it intrinsic} heavy flavour quark 
          components in the proton can be seen by the forward production of the open heavy flavour 
          in $pp$-collisions.
      
          Our calculations of the charmed meson production in $pp$ collisions 
          within the hard scattering model using the PYTHIA8 MC generator \cite{LBPZ:2012} and the 
          PDF including the intrinsic charm  \cite{Nadolsky:2008zw} showed the following.
          We found that the contribution of the {\it intrinsic} charm in the proton could be
          studied in the production of $D$-mesons in $pp$ collisions at the LHC. The IC contribution
          for the single $D^0$-meson production can be large, about 100 $\%$ at high 
          rapidities 3 $\leq y\leq$ 4.5 and large transverse momenta 10 $\leq p_t\leq$ 25 GeV$/$c. 
          For the double $D^0$ production this contribution is not larger than 30 $\%$ at 
          $p_t\geq$ 5 GeV$/$c and  3 $\leq y\leq$ 4.5. These IC contributions for the single and 
          double $D$-meson production \cite{LBPZ:2012} were obtained with the probability of the 
          {\it intrinsic} charm taken to be $w_{c{\bar c}}=$3.5 $\%$ \cite{Nadolsky:2008zw}, and they will 
          decrease by a factor of 3 when $w_{c{\bar c}}\simeq$ 1 $\%$. Therefore, this value can be verified 
          experimentally at the LHCb. 

          The presented predictions could stimulate measurement of the single and double
          D-meson production in $pp$ collisions at the CERN LHCb experiment  in the kinematic region 
          mentioned above to observe a possible signal for the {\it intrinsic} charm. The {\it intrinsic} beauty
          in the proton is suppressed by a factor of 10, therefore its signal in the inclusive spectra
          of $B$-mesons will probably be very weak.

          We also analyzed the inclusive $K^-$-meson production in $pp$ collision at the initial energy 
          $E_p=$158 GeV and gave some predictions for the NA61 experiment going on at CERN, and made some
          predictions for the inclusive spectra of $K^\pm$-mesons produced in $pp$ at the LHC energy $\sqrt{s}=$
          7 TeV. We showed that
          in the inclusive spectrum of $K^-$-mesons as a function of $p_t$ at the initial energy $E_p=$158 GeV and
          some values of their rapidities
          the signal of the {\it intrinsic} strangeness can be visible and reach about 200$\%$ and more at large 
          momentum transfer we took. The signal of the {\it intrinsic} strangeness in the inclusive spectrum
          of $K^-$-mesons produced in $pp$ collision at $\sqrt{s}=$7 TeV can be about 200$\%$ and even
          400$\%$ at $p_t>$15 GeV$/$c, $y=$4.5.     
          The probability of the {\it intrinsic} strangeness to be about 2.5$\%$, as 
          was found from the best description of the HERA and HERMES data on the DIS, see \cite{IS:2012} and
          references therein. 
\section{Acknowledgements}
We are very grateful to A.F Pikelner 
for his help with the MC calculations.
We thank S.J.Brodsky,  M. Gazdzicki, S.M. Pulawski and  A.Rustamov
for extremely helpful discussions and recommendations for the 
predictions on the search for the possible intrinsic heavy flavour components in
$pp$ collisions at high energies.
We are also grateful to V.A.Bednyakov, I.Belyaev, V.Gligorov, H.Jung, B.Kniehl, 
B.Z.Kopeliovich, A.Likhoded, P.Spradlin, M. Poghosyan, V.V Uzhinsky and N.P.Zotov 
for very helpful discussions. 
This work was supported in part by the Russian Foundation for Basic Research, 
grant No: 11-02-01538-a.

\begin{footnotesize}

\end{footnotesize}


\end{document}